\documentclass[twocolumn,floatfix,superscriptaddress,a4paper,showpacs,showkeys,nofootinbib,notitlepage]{revtex4-1}
\usepackage[colorlinks=true,linktocpage=true,linkcolor=blue,citecolor=blue,allcolors=blue]{hyperref}
\usepackage{epsfig}
\usepackage{latexsym}
\usepackage[utf8]{inputenc}
\usepackage{xspace}
\usepackage{indentfirst}
\usepackage{enumerate}
\usepackage{color}

\newenvironment{referee}{\par\color{blue}}{\par}

\graphicspath{{newfigs/},{NA61/},{figs/},{WIP/}}
\usepackage{amsmath}
\usepackage{amssymb}
\usepackage[english]{babel}
\usepackage{url}
\usepackage[normalem]{ulem}

\newcommand{\mean}[1]{\langle #1 \rangle}

\newcommand{\Eq}[1]{Eq.~(\ref{#1})}

\newcommand\ddfrac[2]{\frac{\displaystyle #1}{\displaystyle #2}}

\newcommand{\rom}[1]{\uppercase\expandafter{\romannumeral #1\relax}}

\newcommand{\eq}[1]{\begin{align} #1 \end{align}}

\newcommand{\skewn}[0]{S\sigma}
\newcommand{\kurt}[0]{\kappa\sigma^2}

\newcommand{\cum}[1]{\kappa_{#1}}
\newcommand{\cumt}[1]{\tilde{\kappa}_{#1}}

\newcommand{\w}[1]{\widetilde{#1}}

\begin{document}

 \title{
Higher order conserved charge fluctuations
inside the mixed phase 
}

\author{Roman V. Poberezhnyuk}
\affiliation{Bogolyubov Institute for Theoretical Physics, 03680 Kyiv, Ukraine}
\affiliation{Frankfurt Institute for Advanced Studies, Giersch Science Center,
D-60438 Frankfurt am Main, Germany}

\author{Oleh Savchuk}
\affiliation{Frankfurt Institute for Advanced Studies, Giersch Science Center,
D-60438 Frankfurt am Main, Germany}

\author{Mark I. Gorenstein}
\affiliation{Bogolyubov Institute for Theoretical Physics, 03680 Kyiv, Ukraine}

\author{Volodymyr~Vovchenko}
\affiliation{Nuclear Science Division, Lawrence Berkeley National Laboratory, 1 Cyclotron Road, Berkeley, California 94720, USA}

\author{Horst Stoecker}
\affiliation{Frankfurt Institute for Advanced Studies, Giersch Science Center,
D-60438 Frankfurt am Main, Germany}
\affiliation{
Institut f\"ur Theoretische Physik,
Goethe Universit\"at Frankfurt, D-60438 Frankfurt am Main, Germany}
\affiliation{
GSI Helmholtzzentrum f\"ur Schwerionenforschung GmbH, D-64291 Darmstadt, Germany}

\date{\today}

\begin{abstract}

General formulas are presented  for  higher order  cumulants of the conserved charge statistical fluctuations inside the mixed phase.
As a particular example the van der Waals model in the  grand canonical ensemble is used.
The higher order measures of the conserved charge fluctuations up to the hyperkurtosis are calculated in a vicinity of the critical point (CP). The analysis includes both the mixed phase region and the pure phases on the phase diagram. It is shown that even-order fluctuation measures, e.g. scaled variance, kurtosis, and hyperkurtosis,  have only positive values  in the  mixed phase, and go to infinity at the CP. 
For odd-order measures, such as skewness and hyperskewness, the regions of positive and negative values are found near the left and right binodals, respectively.
The obtained results are discussed in a context of the event-by-event fluctuation measurements in heavy-ion collisions.

\end{abstract}


\maketitle

\section{Introduction}

The structure of the QCD phase diagram 
is one of most interesting unsolved problems in physics. 
Statistical fluctuations of conserved charges are regarded to be sensitive probes of the critical behavior in strongly interacting matter 
\cite{Stephanov:1998dy,Stephanov:1999zu,Athanasiou:2010kw,Stephanov:2008qz,Kitazawa:2012at,Vovchenko:2015uda}.
The fluctuations can be quantified in terms of cumulants (susceptibilities) of the conserved charge distribution. 
Without the loss of generality, we will specifically refer to net baryon number $B$ throughout this work. 
Cumulant of order $j$ can be written as follows:
\eq{\label{cumgen}
\kappa_j=
\left[ \frac{\partial ^j}{\partial t^j} {\rm ln}~ \sum_{r=0}^{\infty}\frac{\mean{B^r}}{r!}t^r\right]_{t=0}~, \quad j=1,2,\ldots,
}
where $\langle...\rangle$ denotes the ensemble average.
Cumulants can be expressed in terms of the moments $\langle B^r \rangle$ explicitly~\cite{Comet}
\eq{\label{cum-j}
\kappa_j = \sum_{k=1}^j (-1)^{k-1}(k-1)!
B_{j,k}(\mean{B},\ldots,\mean{B^{j-k+1}}).
}
Here $B_{j,k}$
are partial exponential Bell polynomials.

It useful to consider ratios of cumulants because such quantities are intensive, i.e. they are volume-independent in the thermodynamic limit. 
The most familiar such measures are 
scaled variance $\omega$, 
skewness $S\sigma$, and kurtosis $\kappa\sigma^{2}$ (see e.g. Ref.~\cite{Karsch:2010ck}):
\eq{\label{cumB}
\mean{B}=\kappa_1,~~~ \omega =\frac{\kappa_2}{\kappa_1},~~~  S\sigma=\frac{\kappa_3}{\kappa_2},~~~ \kappa\sigma^{2}=\frac{\kappa_4}{\kappa_2}.
}
Higher order measures such as hyperskewness $\cum{5}/\cum{2}$ and hyperkurtosis $\cum{6}/\cum{2}$ are also used.

The statistical fluctuations are sensitive to presence of a first order phase transition (FOPT).  
The endpoint of the FOPT is the critical point (CP),
where the cumulant ratios exhibit singular behavior. 
The larger the order of a cumulant is, the stronger is its sensitivity to critical phenomena.

At least two FOPTs are relevant for the QCD phase diagram: (i) the nuclear liquid-gas transition at small temperature $T$ and large baryon chemical potential $\mu$ well
 established both
theoretically~\cite{Sauer:1976zzf,Csernai:1986qf,Serot:1984ey,Zimanyi:1990np,Brockmann:1990cn,Elliott:2013pna,Vovchenko:2016rkn,Poberezhnyuk:2018mwt} and experimentally~\cite{Pochodzalla:1995xy,Natowitz:2002nw,Karnaukhov:2003vp}
and (ii) the hypothetical first-order chiral phase transition at finite baryon densities~\cite{Stephanov:1999zu,Hatta:2002sj,Stephanov:2008qz,Stephanov:2011pb}.
Both transitions are expected to influence the baryon number fluctuations. 
Model calculations suggest that the behavior of baryon number cumulants in certain regions of the phase diagram is determined by a  complex interplay of the chiral and liquid-gas phase transitions~\cite{Mukherjee:2016nhb,Motornenko:2019arp}.

Significant attention has been given to the structure of higher order measures of fluctuations of conserved charges at supercritical temperatures and in pure phases~(see e.g. Refs.~\cite{Vovchenko:2015pya,Stephanov:1999zu,Hatta:2002sj,Stephanov:2008qz,Stephanov:2011pb,Mukherjee:2016nhb,Poberezhnyuk:2019pxs,Motornenko:2019arp}). 
On the other hand, less attention has been paid to the mixed phase.
Nevertheless, it is feasible that a system created in relativistic nucleus-nucleus collisions can enter the mixed phase of a FOPT under certain conditions.
This is especially relevant in view of the plans of the HADES collaboration at Helmholtzzentrum für Schwerionenforschung (GSI) to
measure the higher order net-proton and net-charge fluctuations in central 
Au+Au
reactions at collision energies  $E_{\rm lab} = 0.2A-1.0A~{\rm GeV}$ to probe the liquid-gas FOPT region~\cite{Bluhm:2020mpc}.
The freeze-out of the expanding system created in collisions at these energies may well take place in the mixed phase of the nuclear liquid-gas FOPT.

\begin{figure}
    \centering
\includegraphics[width=.49\textwidth]{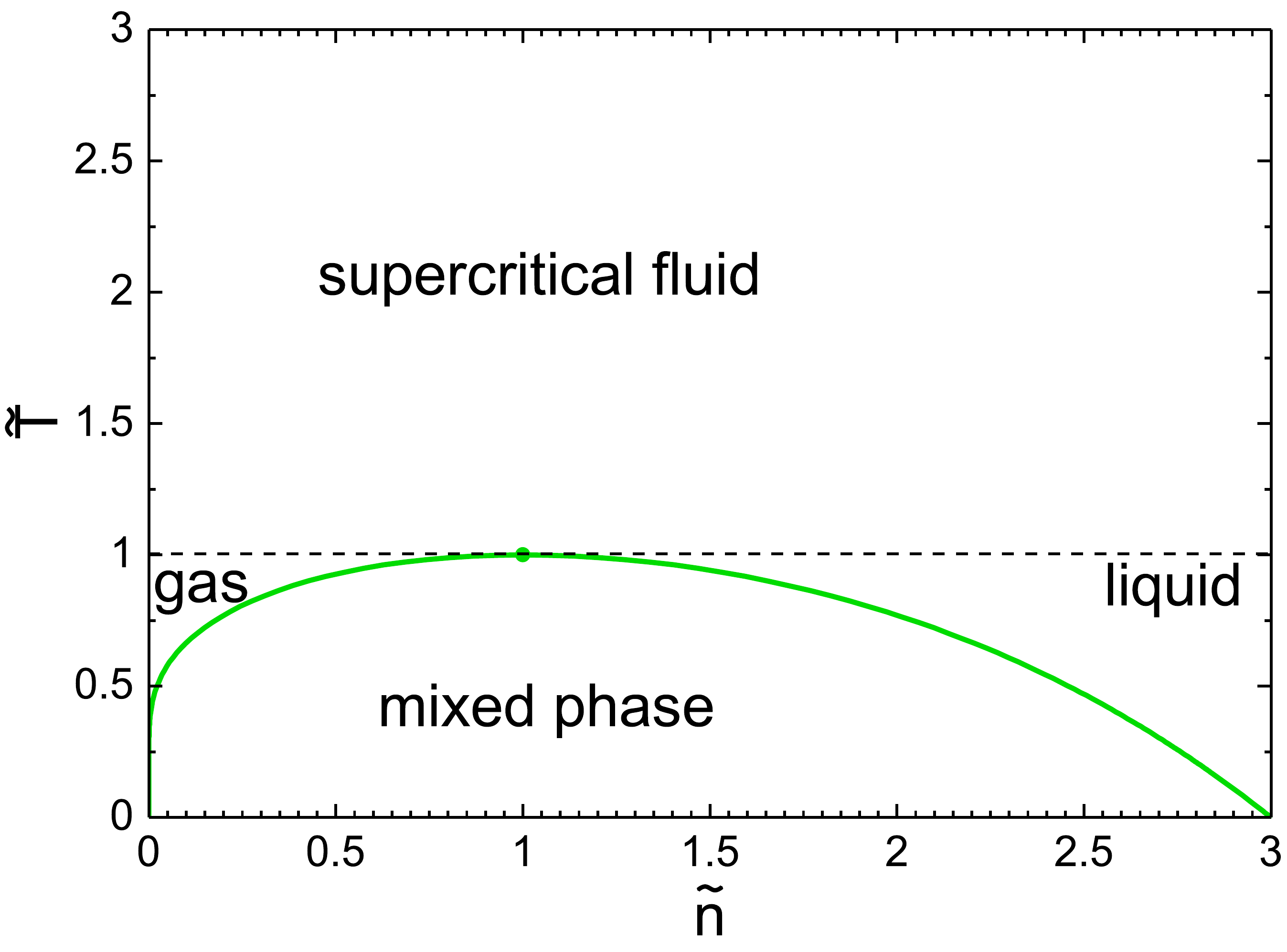}
    \caption{The $(n,T)$ phase diagram of first order phase transition
    in the reduced variables $\widetilde{n}=n/n_c$ and $\widetilde{T}=T/T_c$.
    }
    \label{fig:mp}
\end{figure}

From a theoretical point of view, it is convenient to study the statistical fluctuations using the grand canonical ensemble (GCE).
In the GCE, the cumulants are determined by partial derivatives of the pressure $p$ with respect to a corresponding chemical potential $\mu$:
\eq{\label{pres}
\kappa_j =VT^3~\frac{\partial^j (p/T^4)}{\partial (\mu/T)^ j}~.
}
Here $V$ and $T$ are  the system volume and temperature, respectively.

In $(\mu,T)$ variables, the FOPT is a line at subcritical temperatures~($T<T_c$) that ends at the CP.
Each point on this line corresponds to a coexistence of two phases: a diluted ``gas'' phase with density $n_1$ and a dense ``liquid'' phase with density $n_2$.
The total baryon density $n$ is a superposition of the gaseous and liquid phase densities, and lies anywhere in the range $n \in [n_1,n_2]$.
For this reason it is more appropriate to study the mixed phase phenomena using density-temperature variables $(n,T)$ instead.
Figure~\ref{fig:mp} depicts a typical phase diagram for a system with the liquid-gas FOPT calculated within the van der Waals (vdW) model (see Sec.~\ref{vdw}).
A large fraction of the $(n,T)$ plane at $T<T_c$ corresponds to the mixed phase. 
At each point of the mixed phase the pressures of the first and second phases are equal, $p_1(T,\mu)=p_2(T,\mu)$. 
This is a manifestation of the so-called Gibbs equilibrium condition for the FOPT. 
However, the $T$ and $\mu$ derivatives of the functions $p_1$ and $p_2$ are different. 
Therefore, the statistical fluctuations of conserved charges given by Eq.~(\ref{pres}) differ between the first and second phases.

In this paper, we present a general formalism to calculate the GCE conserved charge cumulants in the mixed phase.
The formalism, presented in Sec.~\ref{formalism}, takes into account the statistical fluctuations in each of the two phases that comprise the mixed phases, as well as fluctuations in the volume fractions occupied by each of the phases. 
The formalism is then applied to describe the behavior of cumulants up to sixth order in the mixed phase of a vdW fluid~(Sec.~\ref{vdw}).
The summary in Sec.~\ref{summary} closes the paper.

\section{Grand-canonical fluctuations in the mixed phase}\label{formalism}

The total system volume $V$ is partitioned in the mixed phase into volumes $V_1 = x V$ and $V = y V$ occupied by the first and second phases, respectively. Here $y \equiv 1-x$.

The $r$th moment of conserved charge distribution is the following:
\eq{\label{mf-moment}
\mean{B^r}=\mean{(B_1+B_2)^r}=V^r \, \mean{(x \rho_1+y \rho_2)^r}~.
}
Here $\rho_1\equiv B_1/V_1$ and $\rho_2=B_2/V_2$ are the baryon densities in the first and second phase, respectively, and $\langle \ldots \rangle$ corresponds to the GCE averaging.
The fluctuating quantities are the densities $\rho_1$, $\rho_2$ and the volume fraction $x$, whereas the total volume $V$ is fixed.
Following Refs.~\cite{Vovchenko:2015vxa,Satarov:2020loq} we assume that
the fluctuations of all these quantities are independent in the thermodynamic limit, i.e. $\mean{\rho_1^l \, \rho_2^m \, x^n} = \mean{\rho_1^l} \, \mean{\rho_2^m} \, \mean{x^n}$ for any non-negative integers $l$, $m$, and $n$.

It is instructive to start with the first moment, $r = 1$.
Equation~(\ref{mf-moment}) in this case reduces to
\eq{\label{mean-B}
\mean{B}=x_0 V n_1 + y_0 V n_2 = V n~,
}
where 
$x_0=\mean{x}$
is the mean volume fraction occupied by the first phase, $y_0\equiv 1-x_0$, and $n_1=\mean{\rho_1}$,  $n_2=\mean{\rho_2}$ are the mean densities in the first and second phases, respectively.
$\mean{B}$ defines the mean baryon density $n$ in the system: $n \equiv \mean{B}/V$. Equation~(\ref{mean-B}) defines $x_0$ in terms of the mean densities:
\eq{\label{x0}
x_0\equiv\mean{x}=\frac{n_2-n}{n_2-n_1}~.
}

To obtain all other cumulants one substitutes Eq.~\eqref{mf-moment} into~\eqref{cum-j}.
The first three cumulants read
\eq{
\kappa_1&=Vx_0n_1+V y_0 n_2=\kappa_{1,1}+\kappa_{1,2}~, \label{k1}\\
\kappa_2&=\kappa_{2,1}\left[1+\frac{\kappa_{2,x}}{x_0^2}\right]+\kappa_{2,2}\left[1+\frac{\kappa_{2,x}}{y_0^2}\right] \nonumber \\
& \quad +(n_2-n_1)^2 V^2\kappa_{2,x}~, \label{k2}\\
\kappa_3&=\kappa_{3,1}\left[1+3\frac{\kappa_{2,x}}{x_0^2}\right]+\kappa_{3,2}\left[1+3\frac{\kappa_{2,x}}{y_0^2}\right]-3(n_2-n_1) \nonumber\\
& \quad \times V\left[\kappa_{2,1}\frac{\kappa_{3,x}+2x_0\kappa_{2,x}}{x_0^2}+
\kappa_{2,2}\frac{\kappa_{3,x}-2y_0\kappa_{2,x}}{y_0^2}\right] \nonumber \\
& \quad +\left[\frac{\kappa_{3,1}}{x_0^3}-\frac{\kappa_{3,2}}{y_0^3}-(n_2-n_1)^3 V^3\right]\kappa_{3,x}~. \label{k3}
}
Here 
$\kappa_{j,1}$~($\kappa_{j,2}$)
is a $j$th-order cumulant
of the $B_1$~($B_2$) fluctuations in the first~(second) phase.  
These cumulants 
describe fluctuations in a pure phase; thus, they should be calculated according to Eq.~(\ref{pres}).  
$\kappa_{j,x}$
is the $j$th order cumulant of the volume fraction $x$ distribution.
It is expressed in terms of cumulants of a subvolume $V_1$ distribution as
$\cum{j,x} \equiv V^{-j}\cum{j}[V_1]$.
In the thermodynamic limit, $V\rightarrow\infty$, all
cumulants of extensive quantities are proportional to the system volume $V$: i.e. $\kappa_{j,1(2)} \sim V$ and $\kappa_j[V_1] \sim V$.
This implies $\cum{j,x} \equiv V^{-j}\cum{j}[V_1] \sim V^{-j+1}$.
Leaving in Eqs.~\eqref{k1}-\eqref{k3} only the terms that are linear in $V$ one obtains the following expressions for the cumulants in the thermodynamic limit:
\eq{
\kappa_1&=\kappa_{1,1}+\kappa_{1,2}~,
\label{kappa-1}\\ 
\kappa_j&=\kappa_{j,1}+\kappa_{j,2}+\left[(n_1-n_2) V\right]^j \, \kappa_{j,x}~,~~~j\ge 2. \label{n-linear} 
}
Cumulants $\kappa_{j,x}$ of the volume fraction parameter $x$ distribution can be expressed in the thermodynamic limit in terms of the GCE cumulants $\kappa_{j,1}$ and $\kappa_{j,2}$ of the two phases. Details of this calculation are given in the Appendix~\ref{app:x}.

Note that the total cumulants reduce to the sum of cumulants of fluctuations in the two phases if the $x$-fluctuations are neglected, i.e. for $x\equiv x_0$ one has
\eq{\label{linear}
\kappa_j &\simeq \kappa_{j,1}+\kappa_{j,2}~,~~~~j=1,2,\ldots.
}
Note that Eq.~\eqref{linear} is also valid away from the thermodynamic limit, i.e. at finite values of the system volume $V$.

It is also instructive to rewrite Eqs.~\eqref{kappa-1} and \eqref{n-linear} in terms of susceptibilities, $\chi_j \equiv \kappa_j / (VT^3)$, which are the intensive measures of particle number fluctuations. One obtains
\eq{
\chi_1 &= x_0 \, \chi_{1,1} + (1-x_0) \, \chi_{1,2}~,
\label{kappa-1-susc}\\ 
\chi_j &= x_0 \, \chi_{j,1}+(1-x_0) \, \chi_{j,2}\nonumber \\ 
& \quad +\frac{\left[(n_1-n_2) V\right]^j}{V\,T^3} \, \kappa_{j,x}~,~~~j\ge 2. \label{n-linear-susc} 
}
The susceptibility $\chi_j$ of particle number fluctuations in the mixed phase corresponds to a linear combination of the pure phase susceptibilities at the left and right binodals plus the contribution from the $x$-fluctuations.

Equations~\eqref{kappa-1} and \eqref{n-linear}~[as well as \eqref{kappa-1-susc}, \eqref{n-linear-susc}] are model-independent expressions describing the GCE fluctuations of a conserved charge in the mixed phase of a FOPT
in the thermodynamic limit.
Model dependence will enter only through explicit form of the cumulants $\kappa_{j,1}$ and $\kappa_{j,2}$.

\section{
van der Waals fluid}
\label{vdw}

In this section we illustrate the general formalism introduced in the previous section
using the vdW equation of state describing Maxwell-Boltzmann interacting particles.
Here we neglect the antiparticles, therefore, the number of particles  plays the role of the conserved charge.

The system pressure of a vdW fluid in a pure phase reads
\eq{\label{pressure}
p(n,T) = \frac{n\,T}{1-b\,n}-an^2~ ,
}
where $a>0$ and $b>0$ are the model parameters describing the attractive and repulsive interactions, respectively.
The CP is defined by conditions~\cite{GNS,LL} 
\eq{\label{cp}
\left(\frac{\partial p}{\partial n}\right)_T=0~,~~~~\left(
\frac{\partial^2 p}{\partial n^2}\right)_T=0~,
}
which give  
\eq{\label{cp-1}
T_c= \frac{8a}{27 b}~,~~~~ n_c=\frac{1}{3b}~,~~~~p_c=\frac{a}{27b^2}~.
}
Introducing reduced variables $\widetilde{T}=T/T_c$, $\widetilde{n}=n/n_c$, and  $\widetilde{p}=p/p_c$ one can rewrite the vdW equation (\ref{pressure}) in a universal form
\eq{\label{vdW-p}
\left(\widetilde{p}~+~3\,\widetilde{n}^2 \right)\,\left(\frac{3}{\widetilde{n}}~-~1\right)=~\widetilde{T}~,
}
which is independent of the specific numerical values of the interaction parameters $a$ and $b$.
This is a particular case of the principle of the corresponding states (see, e.g. Ref.~\cite{GNS}). The phase diagram of the vdW fluid in the $(n,T)$ plane is presented in Fig.~\ref{fig:mp}.

\begin{figure}
\includegraphics[width=.49\textwidth]{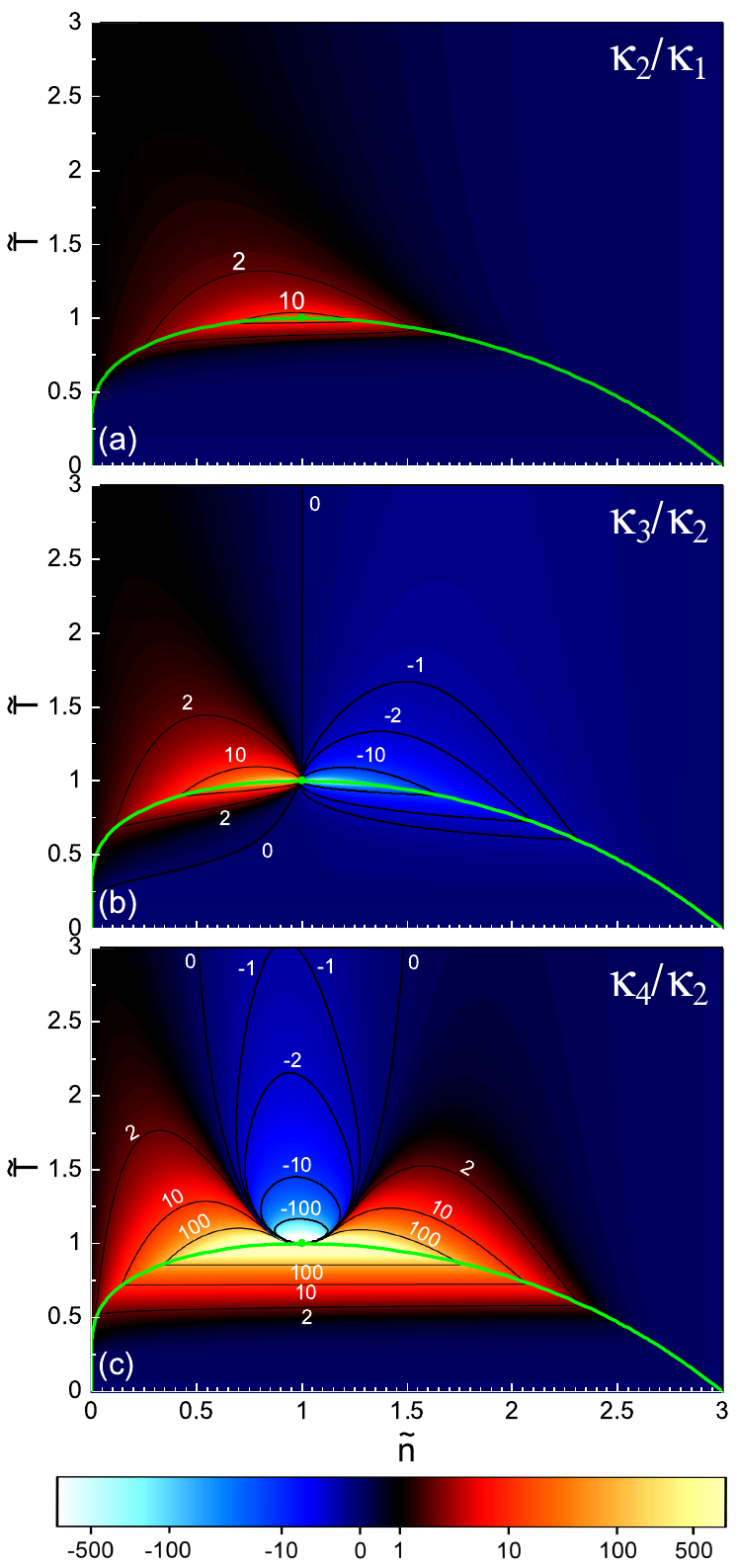}
\caption{\label{general}
The vdW model results in the reduced ($\w{n},\w{T}$) coordinates for the (a) scaled variance, (b) skewness, and (c) kurtosis. 
The mixed phase fluctuation values 
are obtained 
using Eq.~(\ref{linear}). 
The binodals and the CP are represented by the green lines and the green points, respectively.
}
 \end{figure} 

\begin{figure*}
\includegraphics[width=\textwidth]{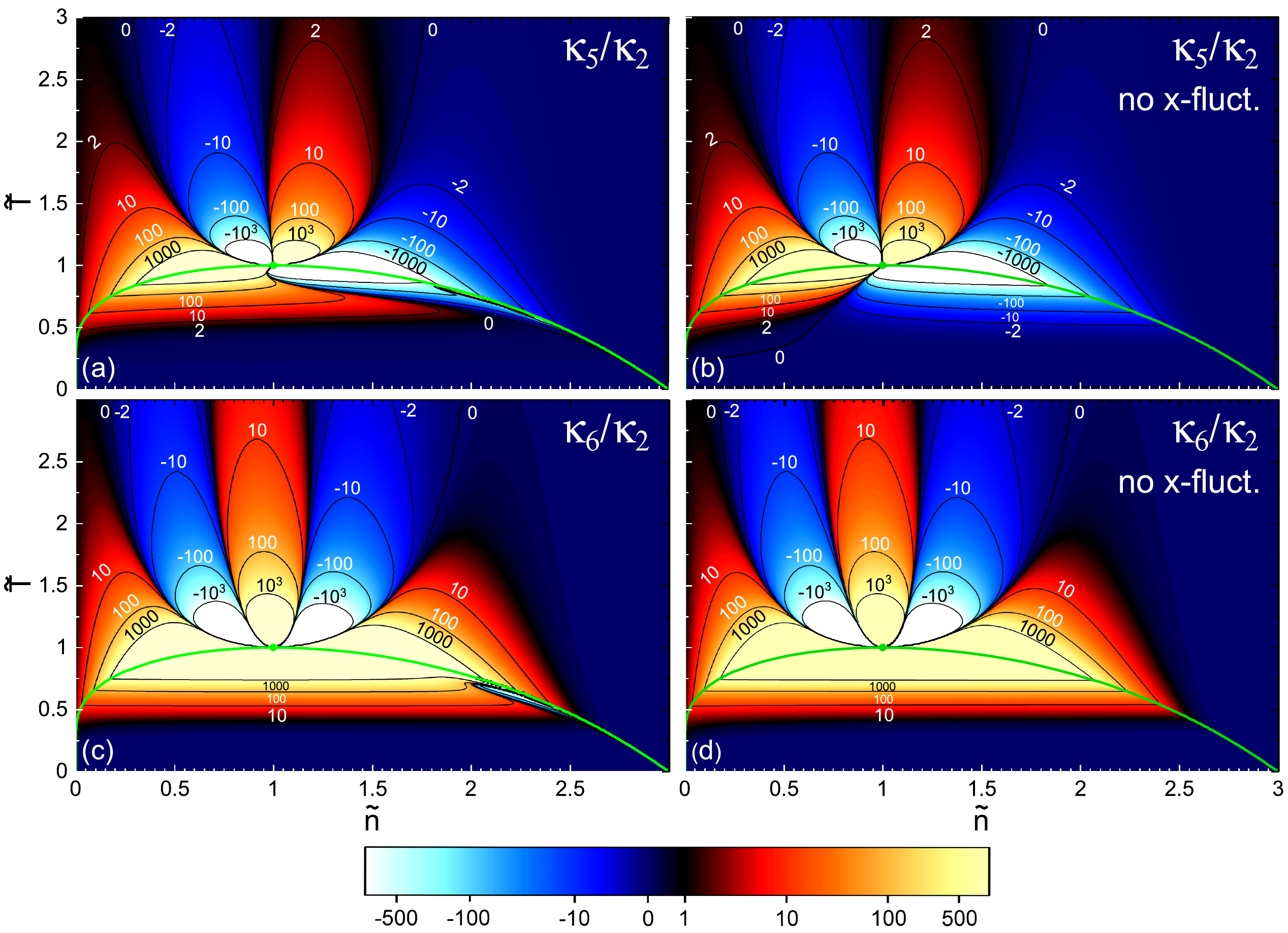}
\caption{\label{fig:hyper}
Hyperskewness and hyperkurtosis
calculated including the $x$-fluctuation effects using Eq.~(\ref{n-linear}) are shown, respectively, in panels (a) and (c).  For comparison the same quantities calculated with Eq.~(\ref{linear}), i.e., when $x$-fluctuations are neglected, are presented in panels (b) and (d). 
}
 \end{figure*}

In the GCE, the vdW model particle number density can be written as follows~\cite{Vovchenko:2015xja}:
\eq{\label{n-vdw}
\widetilde{n} = b~n_{\rm id}(T,\mu)~(3 - \widetilde{n})~\exp\left[-\ddfrac{\widetilde{n}}{3-\widetilde{n}}+\frac{9 \widetilde{n}}{4 \widetilde{T}}\right]~,
}
where $n_{\rm id}(T,\mu)$ is the ideal gas density in the GCE.

The cumulants of vdW model particle number distribution in the GCE can be calculated up to a desired order by iteratively differentiating Eq.~\eqref{n-vdw} with respect to the chemical potential $\mu$~(see Ref.~\cite{Savchuk:2019yxl} for the technical details).
The resulting expressions for the scaled variance,
skewness,
and kurtosis 
for the case of pure phases
read~\cite{Vovchenko:2015xja,Vovchenko:2015uda}:
\eq{
\label{omega-gce}
\omega &
~=~\frac{1}{9}\left[\frac{1}{(3-\widetilde{n})^2}-\frac{\widetilde{n}}
{4 \widetilde{T}}\right]^{-1}~, \\
\label{S-gce}
S\sigma
&~=~ 
\frac{1}{3}\left[\frac{1}{(3-\widetilde{n})^2}-\frac{\widetilde{n}}{4 \widetilde{T}}\right]^{-2}~
\left[\frac{1-\widetilde{n}}{(3-\widetilde{n})^3}\right]~,
\\
\label{kurt-gce}
\kappa \sigma^2
&~=~
3\, (S\sigma)^2 - 2\, \omega \, S\sigma-
54\, \omega^3\frac{\widetilde{n}^2}{(3-\widetilde{n})^4}~.
}
Following the same procedure we also calculate the GCE hyperskewness $\cum{5}/\cum{2}$ and hyperkurtosis $\cum{6}/\cum{2}$.
As the resulting expressions are very lengthy, we do not list them here.
Note that Eqs. (\ref{omega-gce})-(\ref{kurt-gce}) 
describe cumulants of the GCE particle number distribution in pure phases, i.e. at all densities at $T>T_c$ and outside the mixed phase region at $T<T_c$, as well as in metastable phases.
Calculation of fluctuations in the mixed phase, however, requires the use of Eqs.~\eqref{kappa-1} and \eqref{n-linear}.

The boundaries of the mixed phase---the left and right binodals $\widetilde{n}_1(\w{T})$ and $\widetilde{n}_2(\w{T})$---are defined by the Gibbs equilibrium conditions: $\mu(\w{T},\widetilde{n}_1)=\mu(\w{T},\widetilde{n}_2)$ and $p(\w{T},\widetilde{n}_1)=p(\w{T},\widetilde{n}_2)$. 
In the case of vdW model, these conditions lead to 
\eq{
\w{n}_1+\w{n}_2&=\frac{8\w{T}}{(3-\w{n}_1)(3-\w{n}_2)}~,\\\nonumber
\frac{9}{4}\frac{\w{n}_2-\w{n}_1}{ \w{T}}&=
{\rm ln}\left[\frac{3-\w{n}_1}{\w{n}_1}\right]-{\rm ln}\left[\frac{3-\w{n}_2}{\w{n}_2}\right]~~~~~~~~~~~~~~\\
& \quad +\frac{\w{n}_2}{3-\w{n}_2}-\frac{\w{n}_1}{3-\w{n}_1}~.
}
The binodals are depicted in Figs.~\ref{fig:mp}-\ref{fig:hyper} by green lines. 
The fluctuations inside the mixed phase are  presented in Figs.~\ref{general} and \ref{fig:hyper}.  The quantities shown in Figs.~\ref{general},  \ref{fig:hyper}(b), and \ref{fig:hyper}(d) are calculated using  
Eq.~(\ref{linear}), i.e., the $x$-fluctuations are neglected.
The results presented in Figs.~\ref{fig:hyper}(a) and \ref{fig:hyper}(c) are obtained via Eq.~(\ref{n-linear}), where the $x$-fluctuations are taken into account.

The calculations show that the $x$-fluctuations do not affect $\omega$, $\skewn$, and $\kurt$ significantly.
The only exception is a close proximity to the right binodal, where $x\ll 1$ (see Appendix \ref{app:x}).
The differences between Eqs.~(\ref{linear})  and (\ref{n-linear}) would be barely visible in Fig.~\ref{general}. Thus, $\omega$, $\skewn$, and $\kurt$ calculated using Eq.~(\ref{n-linear}) are not presented in Fig.~\ref{general}.

As seen from Fig.~\ref{general}(a),  the scaled variance $\omega\rightarrow +\infty$ at the CP, both inside and outside the mixed phase. This is in agreement with Ref.~\cite{Vovchenko:2015vxa}.
A behavior of $\skewn$ shown in Fig.~\ref{general}(b) is also similar inside and outside the mixed phase. 
However, the behavior of $\kurt$ inside and outside the mixed phase differs drastically. This is shown in Fig.~\ref{general}(c). 
Outside the mixed phase, $\kurt$ attains large positive or negative values in vicinity of the CP, it approaches $+\infty$, $-\infty$, or $0$ at $T\rightarrow T_c$, $n\rightarrow n_c$, i.e., its value depends on the path of approach toward the CP in the $(n,T)$ plane.  
Inside the mixed phase, $\kurt$ is only positive, with large values attained in the vicinity of the CP. 
$\kurt$ tends to $+\infty$ when approaching the CP from inside the mixed phase.
Negative values of $\kurt$ are observed at $T>T_c$ only.
At $T<T_c$ the values of $\kurt$ are only positive, this is a reflection of the fact that $\kurt$ in pure phases outside the mixed phase is positive at subcritical temperatures~\cite{Vovchenko:2015vxa}. 
It thus follows from Eq.~\eqref{linear} that $\kurt$ in the mixed phase is positive as well.
Large negative values of $\kurt$ near the CP are only possible in a small region outside the mixed phase at supercritical temperatures $T \gtrsim T_c$ The same qualitative structure of kurtosis in pure phases around the CP is also present in the Ising model~\cite{Stephanov:2011pb}. 
Our conclusions regarding the behavior of $\kurt$ in and around the mixed phase region near the CP thus applies to the Ising model as well.

Similar arguments are applicable for the hyperkurtosis shown in Figs.~\ref{fig:hyper}(c) and \ref{fig:hyper}(d)  and higher fluctuation measures of even order.
For instance, the structure of hyperkurtosis is more involved in comparison with $\kurt$. The band of large positive values at $n\approx n_c$ is surrounded by two bands of large negative values.
However, close to binodals $\cum{6}/\cum{2}$ becomes positive again (in this aspect $\cum{6}/\cum{2}$ is similar to $\kurt$).
As a consequence, the hyperkurtosis is positive in the whole mixed phase region in accordance with \Eq{linear}.
Thus, we conclude that hyperkurtosis is mostly positive in the vicinity of the CP if the mixed phase region is taken into account.
Negative values of $\cum{6}/\cum{2}$ appear within two relatively narrow bands at supercritical temperatures, $T>T_c$, on the $(n,T)$ phase diagram.

In general, the  structure of the fluctuation measures outside the mixed phase becomes increasingly complex with an increase of their order. 
This is not the case inside the mixed phase region.
Only positive values of the even-order measures are found in the vicinity of the CP inside the mixed phase.
For the odd-order fluctuation measures, e.g. hyperskewness, the mixed phase is split into two regions: (i) the first region has positive values and generally corresponds to lower densities close to the left (gaseous) binodal and (ii) the second region with negative values of odd order cumulants at higher densities close to the right (liquid) binodal.
This is shown for the case of hyperskewness in Figs.~\ref{fig:hyper}(a) and \ref{fig:hyper}(b).
Therefore, the odd-order fluctuation measures may attain both positive or negative values in the mixed phase region, as opposed to even-order cumulants which are predominantly positive inside the mixed phase.
Note, however, that a large positive (hyper)skewness generally corresponds to the phase diagram point close to the gaseous phase~(the left binodal) whereas a large negative one is indicative of the vicinity of the liquid phase (the right binodal).

The $x$-fluctuations do become increasingly important for higher order fluctuation measures. 
The behavior of the hyperskewness $\cum{5}/\cum{2}$ and hyperkurtosis $\cum{6}/\cum{2}$ with account of the $x$-fluctuations are presented in Figs.~\ref{fig:hyper}(a) and \ref{fig:hyper}(c). 
These should be compared to Figs. \ref{fig:hyper}(b) and \ref{fig:hyper}(d), which exhibit the same quantities calculated via Eq.~(\ref{linear}), i.e., accounting for the $x$-fluctuations. 
One sees that $x$-fluctuation effects are only relevant 
near the right binodal.
No qualitative changes in a behavior of the considered fluctuation measures due to a presence of the $x$-fluctuations are found.

\section{Summary}
\label{summary}

In the present work we determined the thermal~(grand-canonical) fluctuations of a conserved charge inside the mixed phase of a first-order phase transition.
As opposed to fluctuations in pure phases, where they are determined solely by the equilibrium properties of that single phase, in the mixed phase the cumulants receive contributions from fluctuations in both the gaseous and the liquid phases, as well from the fluctuations in the volume fractions occupied by the two phases.
Our main result here is given by Eqs.~\eqref{kappa-1} and \eqref{n-linear}, which express the grand-canonical conserved charge cumulants for a point inside the phase coexistence region in the thermodynamic limit for \emph{any} equation of state with a first-order phase transition.

A mixed phase cumulant $\kappa_j$ of order $j$ reduces to a sum of the corresponding pure phase cumulants $\kappa_{j,1}$ and $\kappa_{j,2}$ from each of the two phases plus a contribution from the fluctuations of the relative volume fraction $x$ occupied by the gaseous phase, the cumulants of the $x$-distribution are denoted as $\kappa_{j,x}$.
The cumulants $\kappa_{j,1(2)}$  should be calculated 
in the standard way---as derivatives of the grand potential with respect to the chemical potential on the left~(right) side of the phase coexistence line in the $(\mu,T)$ plane.
The $x$-fluctuation cumulants $\kappa_{j,x}$ can be expressed in terms of the cumulants $\kappa_{j,1(2)}$, although their calculation can be quite involved~(see Appendix~\ref{app:x} for explicit results up to fourth order).
We do observe, however, that the effects of the $x$-fluctuations are found to be negligible for cumulants up to fourth order. 
For the fifth and the sixth orders, notable contributions of the $x$-fluctuations appear in a vicinity of the right binodal.\footnote{Note, however, that quantitative effects of $x$-fluctuations are seen at $x_0 \lesssim 1/2$ in the fifth cumulant.}
Therefore, the simple approximate relation~\eqref{linear} can be used in most practical applications.
This also implies that fluctuations in the mixed phase are mainly determined by the intrinsic properties of the two coexisting phases.

To illustrate our results more explicitly, we used the equation of state of a van der Waals fluid. The cumulant ratios of conserved charge fluctuations such as scaled variance $\cum{2}/\cum{1}$, skewness $\cum{3}/\cum{2}$,
kurtosis $\cum{4}/\cum{2}$, hyperskewness $\cum{5}/\cum{2}$, and hyperkurtosis $\cum{6}/\cum{2}$ were calculated both outside and inside the mixed phase region. 
Outside the mixed phase we reproduce the earlier results from the literature~\cite{Asakawa:2009aj,Stephanov:2011pb,Chen:2015dra,Vovchenko:2015vxa}, where cumulant ratios exhibit increasingly involved structures as the order is increased.
Inside the mixed phase, on the other hand, the structure of cumulants is simpler.
The even-order cumulants are predominantly positive while odd-order cumulants can have either sign, generally attaining positive~(negative) values close to the left~(right) binodal. 
In particular, we conclude that negative values of the kurtosis are consistent with a crossover region just above the critical point, as discussed in Ref.~\cite{Stephanov:2011pb}, but not with any region inside the mixed phase of a FOPT.

The obtained results are relevant in the context of heavy-ion collisions, which create a strongly interacting fluid that may pass through a mixed phase of a FOPT at finite baryon density.
Such a scenario can be probed by event-by-event fluctuation measurements.
In particular, one can consider the well-established nuclear liquid-gas phase transition (LGPT).
A beam energy scan of different colliding ions in a sub-GeV collision energy regime will probe the phase diagram in the vicinity of the nuclear LGPT.  
Such a program can be performed by the HADES experiment at GSI.
If the specific nonmonotonic behavior of high-order cumulants, as calculated here in the grand-canonical limit, is observed in measurements of higher-order fluctuations, this
can be interpreted as a signal of the proximity to the CP.

We note, however, that the results are not directly suitable for quantitative conclusions.
The measurements in heavy-ion collisions are performed in momentum rather than coordinate space. 
Also, even in the scenario where thermalization of fluctuations is achieved in heavy-ion collisions, the measurements will still be affected by global charge conservation, finite system size, and volume fluctuation effects, as has been discussed
in the literature~\cite{Jeon:2000wg,Gorenstein:2011vq,Skokov:2012ds,Bzdak:2012an}.
To address the effects of global conservation in pure phases, a subensemble acceptance method has recently been introduced by us in Refs.~\cite{Vovchenko:2020tsr,Vovchenko:2020gne,Poberezhnyuk:2020ayn}. 
In the future we plan extend this method to address global charge conservation influence on conserved charge fluctuations in the mixed phase.
Another interesting avenue is the so-called strongly intensive fluctuation measures~\cite{Gorenstein:2011vq,Sangaline:2015bma}. 
These quantities are designed to cancel out the geometric effects of the total volume fluctuations, but they are expected to be sensitive to the critical point of a FOPT~\cite{Vovchenko:2016dix}, thus it is of interest to elucidate their behavior in the mixed phase.

\section*{Acknowledgments}

The authors thank Leonid Satarov and Jan Steinheimer for fruitful discussions.
R.P. acknowledge the generous support by the Stiftung Polytechnische Gesellschaft Frankfurt.
V.V. was supported by the
Feodor Lynen program of the Alexander von Humboldt
foundation and the U.S. Department of Energy, 
Office of Science, Office of Nuclear Physics, under contract number 
DE-AC02-05CH11231231. The work of M.I.G. is supported by the Program of Fundamental Research of the Department of Physics and Astronomy of National Academy of Sciences of Ukraine.
H.St. appreciates the Judah M. Eisenberg Professur Laureatus for Theoretical Physics of the Walter Greiner Gesellschaft zur Foerderung der physikalischen Grundlagenforschung at the FB Physik of Goethe Universitaet Frankfurt.
 This work was supported by the DAAD through a PPP exchange grant. Computational resources were provided by the Frankfurt Center for Scientific Computing (Goethe-HLR).
 
\begin{widetext}

\appendix

\section{Evaluation of $x$ fluctuations}
\label{app:x}

As stated in Sec.~\ref{formalism}, we assume that in thermodynamic limit, i.e. for $V \to \infty$, correlations between baryon numbers $B_1$,$B_2$ and $x$ can be neglected.
Thus, in the following calculations we use the canonical ensemble with  $B_1,B_2$ fixed to their mean values $B_1=\mean{B_1}$, $B_2=\mean{B_2}$.
Furthermore, in thermodynamic limit the surface terms between coexistent phases can be neglected and, in similarity with Refs.~\cite{Vovchenko:2020tsr,Poberezhnyuk:2020ayn}, the partition function can be presented as a product of partition functions of the two subsystems.
Then the probability $P(x)$ is proportional to the product of the canonical partition functions of the first and second phase:
\eq{
P(x) &  \propto Z
(T,x V,\mean{B_1}) \, Z
(T,y V,\mean{B_2}). \label{PB}
}
Here $y \equiv 1 -x$, $\mean{B_1}=V x_0 n_1$, $\mean{B_2}=V y_0 n_2$, and $x_0$ is given by \Eq{x0}. 

In the thermodynamic limit, the above results can be generalized, since in this case 
the canonical partition function can be 
expressed through the volume-independent free-energy density $f$: $Z(T,V,B) = \exp\left[-\frac{V}{T} \, f(T,n) \right]$ with $n \equiv B/V$ being the conserved baryon density.
Thus,
\eq{
P(x)   \propto \exp\left[-~V~
\frac{
x f\left(T,\ddfrac{x_0}{x}n_1\right)+
y f\left(T,\ddfrac{y_0}{y}n_2\right)
}{T}\right].
}
To evaluate $\cum{j,x}$ we introduce the cumulant generating function $\psi_{x}(t)$:
\eq{
\psi_{x}(t) & \equiv \ln \mean{e^{t\,x}} =\ln \int dx~ e^{t\,x} P(x)
= \ln \left\{ \int dx~ \, \exp\left[ t\,x -V~
\frac{
x f\left(T,\ddfrac{x_0}{x}n_1\right)+
y f\left(T,\ddfrac{y_0}{y}n_2\right)
}{T} \right]  \right\} + \tilde{C}.
}
Here 
$\tilde{C}$ is
an irrelevant normalization constant. The cumulants, $\cum{j,x}$, correspond to the Taylor coefficients of $\psi_{x}(t)$:
\eq{
\cum{j,x} = \left. \frac{\partial^j \psi_{x}(t)}{\partial t^j} \right|_{t=0} \equiv \left.  \cumt{j,x}(t) \right|_{t=0}.
}
Here we have introduced a shorthand,  $\cumt{j,x}(t)$, for the $n$th derivative of the generating
function at arbitrary values of $t$, which we subsequently refer to as $t$-dependent
cumulants. Clearly, all higher order cumulants are given as a $t$-derivative of the first order
$t$-dependent cumulant, $\cumt{1,x}(t)$, which is given by 
\begin{align}
\cumt{1,x}(t)= \frac{\partial\psi_{x}(t)}{\partial t}=  \frac{\int_0^1 dx~ x \, \tilde{P}(x;t)}{\int_0^1 dx~ \tilde{P}(x;t)} = \mean{x(t)}
\end{align}
with the (un-normalized) $t$-dependent probability
\eq{
\tilde{P}(x)   \propto \exp\left[t\,x -V~
\frac{
x f\left(T,\ddfrac{x_0}{x}n_1\right)+
y f\left(T,\ddfrac{y_0}{y}n_2\right)
}{T}\right].
}

One can check that in the thermodynamic limit, $V \to \infty$, $\tilde{P}$ 
has a sharp maximum at the mean value
of $x$, $\mean{x(t)}$.
The condition $\partial \tilde{P}(x;t) / \partial x = 0$ determines the location of this maximum
resulting in an implicit relation that determines $\mean{x(t)}$:
\eq{\label{eq:Q1t}
t = \frac{V}{T}\left[f\left(T,\ddfrac{x_0}{x}n_1\right)-
f\left(T,\ddfrac{y_0}{y}n_2\right)-\ddfrac{x_0}{x}n_1\mu\left(T,\ddfrac{x_0}{x}n_1\right)+\ddfrac{y_0}{y}n_2\mu\left(T,\ddfrac{y_0}{y}n_2\right)\right].
}
Here $x=x(t)$, $n_1=n_1(t) = \mean{B_1} /
(x(t) V)$, $n_2=n_2(t) = \mean{B_2} /
([1-x(t)] V)$. Here we also used a thermodynamic relation 
\eq{\label{td-identity}
\left[\frac{\partial f(T,n)}{\partial n}\right]_T =  \mu(T,n).
}
The solution to Eq.~\eqref{eq:Q1t} at $t = 0$ is $\mean{x(t=0)} = x_0$, as should be by construction.

The second cumulant is determined by the $t$-derivative of $\cumt{1,x}$, i.e. $\cumt{2,x} = \partial \cumt{1,x} / \partial t = \mean{x'(t)}$.
To calculate $\mean{x'(t)}$ we differentiate Eq.~\eqref{eq:Q1t} with respect to $t$.
To evaluate the $t$-derivative of the right-hand side  of~\eqref{eq:Q1t} we apply the chain rule 
$\partial  \mu / \partial t = [\partial  \mu(T,n) / \partial n]_T \, \, [\partial  n(t) / \partial t],$
$\partial  \cum{i} / \partial t = [\partial  \cum{i}(T,n) / \partial n]_T \, \, [\partial  n(t) / \partial t],$
\eq{
\left[\frac{\partial \mu(T,n)}{\partial n}\right]_T=\frac{TV}{\cum{2}},~~~~~\left[\frac{\partial \cum{i}}{\partial n}\right]_T =\left[\frac{\partial \cum{i}}{\partial \mu}\right]_T\left[\frac{\partial\mu(T,n)}{\partial n}\right]_T =V\frac{\cum{i+1}}{\cum{2}}
}
and use the thermodynamic identity~\eqref{td-identity}.
The solution for the resulting equation for $\mean{x'(t)} \equiv \cumt{2,x}$ at $t = 0$ gives the second order  cumulant:
\eq{\label{k2x}
 \cum{2,x} &=\frac{1}{V^2}~\frac{\cum{2,1}~\cum{2,2}}{n_2^2~\cum{2,1}~+~n_1^2~\cum{2,2}}~.
}
This expression is in agreement with a result of Refs.~\cite{Vovchenko:2015vxa,Satarov:2020loq}, obtained there for, respectively, the vdW  and Skyrme-like scalar interaction equations of state.

To evaluate the higher-order cumulants, $\cum{j,x}$ with $j \geq 3$, we iteratively differentiate the $t$-dependent cumulants $\cumt{j,x}(t)$ with respect to $t$, starting from $\cumt{j,x}(t)$, and make use of the expressions
for $\mean{x^{(j-2)}(t)}$.
The result for third and fourth order cumulants is the following:
\eq{\label{k3x}
\cum{3,x} &=\frac{y_0 n_1^2 \cum{2,2}^3\left(3\cum{2,1}^2-\mean{B_1}\cum{3,1}\right)-x_0 n_2^2\cum{2,1}^3\left(3\cum{2,2}^2-\mean{B_2}\cum{3,2}\right)}{V^4 x_0 y_0 \left(n_2^2\cum{2,1}+n_1^2\cum{2,2}\right)^3}\\\nonumber
\cum{4,x}&=\frac{1}{V^6 x_0^2 y_0^2 (n_2^2\cum{2,1}+n_1^2\cum{2,2})^5}
\left[5 x_0^2 n_2^4
\cum{2,1}^5 \cum{2,2}
(3 \cum{2,2}^2 - 2 \mean{B_2} \cum{3,2}) +
5 y_0^2 n_1^4
\cum{2,2}^5 \cum{2,1}
(3 \cum{2,1}^2 - 2 \mean{B_1} \cum{3,1})
\right.\\\nonumber
& + 2 x_0 y_0 \left(n_1 n_2 \cum{2,1} \cum{2,2} \right)^2 \left(
4 V n_1 \cum{2,2}^2 \cum{3,1} + 4 V n_2 \cum{2,1}^2 \cum{3,2}
+ 5 \mean{B_2} \cum{2,1}^2 \cum{3,2} + 5 \mean{B_1} \cum{2,2}^2 \cum{3,1} - 15 \cum{2,1}^2 \cum{2,2}^2
\right)\\\nonumber
& + \left(V x_0 y_0\right)^2 \left\{ 
n_1^6 \cum{2,2}^5 \cum{4,1} + n_2^6 \cum{2,1}^5 \cum{4,2} + 
n_1^4 n_2^2 \cum{2,2}^4\left(\cum{2,1}\cum{4,1}-3\cum{3,1}^2\right)+
n_2^4 n_1^2 \cum{2,1}^4\left(\cum{2,2}\cum{4,2}-3\cum{3,2}^2\right)
\right.\\\label{k4x}
& \left.\left.- 6 n_1^3 n_2^3 \cum{2,1}^2 \cum{2,2}^2 \cum{3,1} \cum{3,2}\right\} - 12 n_1^2 n_2^2 \cum{2,1}^4 \cum{2,2}^4\right]
}

Equations (\ref{k2x})-(\ref{k4x}) are model-independent, i.e., they are applicable for an arbitrary equation of state in the thermodynamic limit.  
The expressions for $\cum{5,x}$, $\cum{6,x}$, and higher cumulants can be obtained using the same logic.
In two limiting cases, namely $x_0\rightarrow 0$ and $x_0\rightarrow 1$, one has either $\cum{j,1}\rightarrow 0$ in the first case and $\cum{j,2}\rightarrow 0$ in the second case, thus, $\cum{j,x}\rightarrow 0$. As a result, cumulants $\cum{j}$ stay continuous as one crosses the binodals. 
The expressions for $\cum{j,x}$ are simplified in a limit $n_1\ll n_2$:
\eq{
(n_2-n_1)^2 V^2 \cum{2,x} &= \cum{2,2}\\
(n_2-n_1)^3 V^3 \cum{3,x} &= \cum{3,2} - 3 \frac{\cum{2,2}^2}{\mean{B_2}}\\
(n_2-n_1)^4 V^4 \cum{4,x} &= \cum{4,2} + 5 \cum{2,2} \frac{3 \cum{2,2}^2 - 2 \mean{B_2}\cum{3,2}}{\mean{B_2}^2}~,
}
where as before $\mean{B_2},\cum{j,2}\sim (1-x)V$.
The condition $n_1\ll n_2$ can be realized e.g. in the low-temperature limit of a liquid-gas transition.
The $x$-fluctuations in this case are proportional to $(1-x)$;
thus, they are mostly relevant in the vicinity of the second binodal, where $x\ll 1$.

\end{widetext}

\bibliography{main.bib}

\end{document}